\documentclass[a4paper,11pt]{article}
\usepackage{jheppub} 
\usepackage{jheppub} 

\usepackage[T1]{fontenc} 
\usepackage{bm}
\usepackage{url}
\DeclareMathAlphabet{\pazocal}{OMS}{zplm}{m}{n}
\usepackage{braket}
\usepackage{amsmath}
\usepackage{amssymb}
\usepackage{leftidx}
\usepackage[colorinlistoftodos]{todonotes}
\usepackage{bm}
\usepackage{slashed}
\usepackage{cancel}
\usepackage{upgreek}
\usepackage{comment}
\usepackage{hyperref}

\newcommand{\diff}{\mathrm{d}}
\newcommand{\vphi}{\varphi}

\title{8D conformal gravity with Einstein sector, and its relation 
to the $Q$-curvature}

\author[a]{Nicolas Boulanger,}
\author[b,c]{Davide Rovere}

\affiliation[a]{Service de Physique de l’Univers, Champs et Gravitation,
Universit\'e de Mons – UMONS, 20 place du Parc, 7000 Mons, Belgium 
\vspace{0.1cm}}
\affiliation[b]{Dipartimento di Fisica e Astronomia ‘G. Galilei’, Università di Padova,
Via Marzolo 8, 35131 Padova, Italy \vspace{0.1cm}}
\affiliation[c]{INFN, Sezione di Padova,
Via Marzolo 8, 35131 Padova, Italy\vspace{0.1cm}}

\emailAdd{nicolas.boulanger@umons.ac.be} 
\emailAdd{davide.rovere@studenti.unipd.it}

\abstract{
We first streamline the construction of the unique six-dimensional 
conformal gravity action found by L{\"u}, Pang and Pope, 
that admits Einstein metrics as solutions to the field equations. 
We then prove that there exists a unique eight-dimensional conformal 
gravity action that admits Einstein metrics as solutions to the 
field equations, and explicitly build the corresponding action.
Finally, we relate these results to Branson's $Q$-curvature 
and the Fefferman-Graham obstruction tensor,  
to conjecture that on every even-dimensional space there exists 
a unique -- up to boundary terms -- conformally-invariant gravity 
theory that is extremised by Einstein metrics.}

\begin{document}
\maketitle

\section{Introduction}
Conformal gravity is a privileged model for an extension of 
Einstein's general relativity theory, since on top of the usual 
diffeomorphism symmetries, it is invariant under Weyl rescalings 
of the metric. 
As a classical theory of gravity, four-dimensional conformal gravity, 
also called  Weyl gravity, was investigated in great details by several 
authors, see in particular 
\cite{Stelle:1977ry, Riegert:1984zz, Mannheim:1990ya, Klemm:1998kf, Dzhunushaliev:1999fy, Emparan:2001wk, Flanagan:2006ra, Oliva:2012zs, Lu:2012xu, Lu:2015cqa}. 
A review of the relevance of Weyl gravity throughout the last 
decades can be found in \cite{Scholz:2017pfo} to which we refer 
for more references. 

Viewed as a quantum theory, conformal gravity was 
shown by Stelle \cite{Stelle:1976gc} to be renormalizable, 
albeit non-unitary.
This triggered an important body of works, where in particular Einstein 
gravity was argued to emerge from quantum corrections to Weyl 
gravity \cite{Adler:1982ri,Zee:1983mj}; for a review, see e.g. 
\cite{Salvio:2018crh}. 
The non-unitarity of conformal gravity has been discussed in many 
references, see e.g. \cite{Fradkin:1985am,Antoniadis:1986tu}, where 
precisely the non-unitary sector can be decoupled from the space 
of solutions to conformal gravity by imposing appropriate boundary 
conditions, thereby leaving the space of solutions of Einstein's 
equations with a cosmological constant \cite{Maldacena:2011mk}.
Indeed, while the solutions to Einstein's equations with 
a cosmological constant -- namely, Einstein manifolds -- 
are also solutions to the field equations of conformal gravity, 
the converse is not true. Nevertheless, as we mentioned above,
Maldacena \cite{Maldacena:2011mk} showed that the  non-Einstein 
metrics of four-dimensional conformal gravity 
can be eliminated by imposing an appropriate Neumann boundary condition 
on the asymptotic (anti) de Sitter (A)dS spacetime metric, which 
constitutes a very interesting and concrete relation between Weyl 
and Einstein gravity theories. 

Whether this connection holds in higher spacetime dimensions $D$ 
was tested in six-dimensional conformal gravity \cite{Anastasiou:2020mik}, 
where the action actually is a two-parameter family of actions, 
due to the fact that in 6$D$ there exist three linearly 
independent scalar densities that are strictly Weyl-invariant 
\cite{Graham1985,parker1987invariants}; see e.g. 
\cite{Erdmenger:1997gy} for a review, and below in the 
body of the paper.
Instead, in four dimensions there is only one Weyl-invariant scalar 
density, the one leading to Weyl gravity. 
It was found in \cite{Lu:2013hx} that, up to an overall factor
in front of the action functional, there is a \emph{unique} linear 
combination of the three local conformal invariants in 6$D$ 
for which Einstein metrics are solutions to the corresponding 
variational problem. In the present technical note, we want to 
see whether this property extends to eight dimensions $D=8$. 

For the construction of the Lagrangian density, one has to start 
from the list of possible Weyl-invariant scalar densities in 8$D$, 
also called local (or pointwise) conformal invariants, which 
were classified in \cite{Boulanger:2004zf} by using the Weyl-covariant
calculus developed in \cite{Boulanger:2004eh}. These purely algebraic 
tools were also used in \cite{Boulanger:2018rxo} to determine the 
general structure of global conformal invariants on manifolds of 
arbitrary dimension.
It was already known from \cite{Alexakis:2005ft,alexakis2012decomposition} 
that, on closed manifolds of even dimensions $D=2m$, global conformal 
invariants are given by the integral over the manifold of the Euler density 
plus a linear combination of the local conformal invariants in that
dimension, plus total derivatives. 
On manifolds of dimension $D=4m-1$, $m\in \mathbb{N}^+\,$, 
further global conformal invariants were found in 
\cite{Boulanger:2018rxo}, thereby completing the results of 
\cite{alexakis2012decomposition}. 

In Section \ref{sec:Weyl-covariant} we briefly review 
the Weyl-conformal calculus developed in \cite{Boulanger:2004eh}.
Then, in Section \ref{sec:four and six} we review the theories of 
conformal gravity in four and six dimensions, that admit an Einstein 
sector. In Section \ref{sec:Q-curvature} we then discuss the notion 
of $Q$-curvature and illustrate it explicitly in dimensions two, 
four and six.
Then, in Section \ref{sec:8D} we construct the most general conformal 
gravity theory in eight dimensions, that admits an Einstein sector, and 
find that the result is unique, up to boundary terms and an additive 
constant proportional to the Euler characteristic.
We relate this action to the eight-dimensional $Q$-curvature, to find 
that, up to boundary terms and an additive constant, our action 
coincide with the (normalised) integrated $Q$-curvature. 
We end the note in Section \ref{sec:Discu} with a general discussion 
of both the $Q$-curvature and the Fefferman-Graham obstruction tensor
in arbitrary even dimension $D=2m$. We conjecture that there is only one 
conformal gravity action in even dimension, that admits an Einstein sector. 
It coincides with the integrated $Q$-curvature, 
up to normalisation, boundary terms, and additive constant proportional 
to the Euler characteristic $\chi(M_{2m})$ of the manifold.

\section{Weyl-covariant tensor calculus}
\label{sec:Weyl-covariant}

The problem of classifying all the Weyl-invariant scalar densities 
built out of a metric in arbitrary (even) dimension is famously
difficult, see e.g. \cite{Graham1985,bailey1994invariant,bailey1994thomas,gover2001invariant,gover2003standard,graham2005ambient} 
and refs. therein. 
The problem is very simple in four dimensions for which the square 
of the Weyl tensor gives the solution, whereas it is already much 
more complicated in six dimensions \cite{parker1987invariants}.
In eight dimensions, the classification of the Weyl-invariant scalar 
densities built out of a metric tensor was obtained in 
\cite{Boulanger:2004zf}. This classification relies on the 
Weyl-covariant tensor calculus developed in \cite{Boulanger:2004eh}
that we will briefly review in this section, as it is also 
instrumental in the classification of the Weyl-invariant action
functionals that admit Einstein metrics as solutions to the variational 
problem. We use the conventions and notation of 
\cite{Boulanger:2007ab,Boulanger:2007st},
where the classification of Weyl anomalies in arbitrary dimension 
was obtained.

First of all, we recall that the Weyl tensor is the traceless part 
of the Riemann curvature tensor. In components, we have 
\begin{equation}
W^\mu{}_{\nu\rho\sigma} = R^\mu{}_{\nu\rho\sigma} 
     - 2\left( \delta^{\mu}{}_{[\rho}K_{\sigma]\nu} 
     - g_{\nu[\rho} K_{\sigma]}{}^\mu\right) \;,
\end{equation}
in terms of the components of the Riemann tensor and of the Schouten 
tensor 
\begin{equation}
K_{\mu\nu} = \frac{1}{D-2}\left(R_{\mu\nu} - 
     \frac{1}{2(D-1)} g_{\mu\nu}\right)\;.
\end{equation}
Under infinitesimal Weyl rescalings of the metric
\begin{equation}
    \delta_\sigma g_{\mu\nu} = 2\,\sigma(x)\,g_{\mu\nu}\;,
\end{equation}
the components of Weyl tensors are invariant:  
$\delta_\sigma W^\mu{}_{\nu\alpha\beta} = 0\,$.
Denoting by $\Delta_\mu{}^\nu\,$ the $GL(D)$ generators that act on tensors 
through $\Delta_\mu{}^\nu \,T^\alpha_\beta = \delta^\nu_\beta T^\alpha_\mu - 
\delta^\alpha_\mu T^\nu_\beta\,$, 
the symbol $\nabla_\mu = \partial_\mu - \Gamma_{\mu\nu}{}^\rho\,\Delta_\rho{}^\nu$ 
denotes the usual torsion-free metric-compatible 
(Levi-Civita) covariant derivative associated with the Christoﬀel symbols
$\Gamma_{\mu\nu}{}^\rho$, in terms of which 
$R^\mu{}_{\nu\rho\sigma} = \partial_{\rho}\Gamma_{\nu\sigma}{}^{\mu}+\ldots$.
The commutator of covariant derivatives gives  $[\nabla_\mu,\nabla_\nu]V^\rho=R^\rho{}_{\sigma\mu\nu}V^\sigma\,$ 
and, in general, 
$[\nabla_\mu,\nabla_\nu] = R_{\mu\nu\rho}{}^\sigma{}\,\Delta_\sigma{}^\rho\,$.
The components of the Cotton tensor are given by 
$C_{\alpha\rho\sigma} = 2\, \nabla_{[\sigma} K_{\rho]\alpha} \equiv 
\nabla_{\sigma} K_{\rho\alpha} - \nabla_{\rho} K_{\sigma\alpha}\,$.
The Weyl-covariant derivative constructed in \cite{Boulanger:2004eh} 
is given by 
\begin{equation}
{\cal D}_{\mu} = \nabla_{\mu}+K_{\mu\alpha}\,\boldsymbol{\Gamma}^\alpha\;,
\end{equation}
where we refer to this work for the definition of 
the generators $\boldsymbol{\Gamma}^\alpha$; see also below for 
a few examples.  
The important property of the Weyl-covariant 
derivative $\cal D$ is that its curvature vanishes if and only if the 
metric is conformally flat. Explicitly, one has \cite{Boulanger:2004eh}
\begin{equation}
    [{\cal D}_\mu,{\cal D}_\nu] = 
      W_{\mu\nu\rho}{}^\sigma{}\,\Delta_\sigma{}^\rho
    - C_{\alpha\mu\nu}\,\boldsymbol{\Gamma}^\alpha\;.
\end{equation}
The first term on the right-hand side is the same as in the expression 
for the commutator of the Levi-Civita covariant derivative, 
except that now the Weyl tensor replaces the Riemann curvature tensor.
The second term on the right-hand side brings the Cotton tensor, which is 
the conformal field strength in 3D, where the Weyl tensor identically vanishes.
In dimensions $D>3\,$, the Cotton tensor can be written as a covariant divergence 
of the Weyl tensor, viz., 
$C_{\alpha\rho\sigma} = -\frac{1}{D-3}\,\nabla_\mu W^\mu{}_{\alpha\rho\sigma}\,$. 

Similarly to the fact that the tensors in (pseudo)Riemann 
geometry are given by the metric tensor, the Riemann tensor, all its 
covariant derivatives and traces thereof using the (inverse)metric tensor, 
the set of \emph{$W$-tensors} is given by the Weyl tensor, all its Weyl-covariant 
derivatives and their non-trivially vanishing traces. 
We introduce super indices and the notation
\[\{W_{\Omega_0}, W_{\Omega_1}, \ldots, W_{\Omega_k}, \ldots\}
=\{W^\mu{}_{\nu\rho\sigma}, {\cal D}_{\alpha_1} W^\mu{}_{\nu\rho\sigma},
\ldots, {\cal D}_{\alpha_k} {\cal D}_{\alpha_{k-1}}\ldots {\cal D}_{\alpha_1} W^\mu{}_{\nu\rho\sigma},\ldots\}\;.\]
The defining property of the $W$-tensors is that they transform, under infinitesimal 
Weyl rescalings of the metric, 
with the first derivative of the Weyl parameter only \cite{Boulanger:2004eh}:
\begin{equation}
 \delta_\sigma W_{\Omega_i} = \partial_\alpha\sigma \,
 [{T}^{\alpha}]_{\Omega_i}{}^{\Omega_{i-1}}\,W_{\Omega_{i-1}}\;.   
\end{equation}
We will also use the notation 
$W^\mu{}_{\nu\rho\sigma,\alpha_1}:={\cal D}_{\alpha_1} W^\mu{}_{\nu\rho\sigma}\,$, 
$W^\mu{}_{\nu\rho\sigma,\alpha_1\alpha_2}:={\cal D}_{\alpha_2}{\cal D}_{\alpha_1} W^\mu{}_{\nu\rho\sigma}\,$, etc. 

By introducing the tensor 
${\cal P}^{\alpha\nu}_{\mu\beta}:=-g^{\alpha\nu}g_{\mu\beta}
+\delta^\alpha_\mu\delta^\nu_\beta+\delta^\alpha_\beta\delta^\nu_\mu\,$, 
we can present the first few $W$-tensors as follows: 
\begin{align}
W_{\Omega_0} =~& W^\mu{}_{\nu\rho\sigma}\;,
\nonumber \\
W_{\Omega_1} =~& W^\mu{}_{\nu\rho\sigma,\alpha_1} 
= {\cal D}_{\alpha_1}\,W^\mu{}_{\nu\rho\sigma} 
= {\nabla}_{\alpha_1}\,W^\mu{}_{\nu\rho\sigma}\;,
\nonumber \\
W_{\Omega_2}  =~&
\nabla_{\alpha_2}\,W^\mu{}_{\nu\rho\sigma,\alpha_1} 
- K_{\alpha_2 \lambda}\,{\cal P}^{\lambda\delta}_{\epsilon\alpha_1}\,
\Delta_\delta{}^\epsilon\,W^\mu{}_{\nu\rho\sigma}  = 
{\cal D}_{\alpha_2}{\cal D}_{\alpha_1} W^\mu{}_{\nu\rho\sigma}
=W^\mu{}_{\nu\rho\sigma,\alpha_1\alpha_2}\;,
\nonumber \\
W_{\Omega_3} =~& \nabla_{\alpha_3}\,W^\mu{}_{\nu\rho\sigma,\alpha_1\alpha_2} -K_{\lambda\alpha_3}\,(\delta^\gamma_{\alpha_1}\,
{\cal P}^{\lambda\delta}_{\epsilon\alpha_2}\,
\Delta_\delta{}^\epsilon + \delta^\gamma_{\alpha_2}\,
{\cal P}^{\lambda\delta}_{\epsilon\alpha_1}\,\Delta_\delta{}^\epsilon - 
{\cal P}^{\lambda\gamma}_{\alpha_1\alpha_2})\,W^\mu{}_{\nu\rho\sigma,\gamma}\;,
\nonumber\\
W_{\Omega_4} =~&
\nabla_{\alpha_4}\,W^\mu{}_{\nu\rho\sigma,\alpha_1\alpha_2\alpha_3} - K_{\lambda\alpha_4}\,\times
\nonumber\\
& \times(\delta^{\gamma_1}_{\alpha_1}\,\delta^{\gamma_2}_{\alpha_2}\,
{\cal P}^{\lambda\delta}_{\epsilon\alpha_3}\,\Delta_\delta{}^\epsilon + \delta^{\gamma_1}_{\alpha_1}\,\delta^{\gamma_3}_{\alpha_2}\,
{\cal P}^{\lambda\delta}_{\epsilon\alpha_2}\,\Delta_\delta{}^\epsilon 
+ \delta^{\gamma_1}_{\alpha_2}\,\delta^{\gamma_2}_{\alpha_3}\,
{\cal P}^{\lambda\delta}_{\epsilon\alpha_1}\,\Delta_\delta{}^\epsilon \,+
\nonumber\\
& \quad - \delta^{\gamma_1}_{\alpha_2}\,{\cal P}^{\lambda\gamma_2}_{\alpha_1\alpha_3} 
- \delta^{\gamma_1}_{\alpha_1}\,{\cal P}^{\lambda\gamma_2}_{\alpha_2\alpha_3}- 
\delta^{\gamma_1}_{\alpha_3}\,
{\cal P}^{\lambda\gamma_2}_{\alpha_1\alpha_2})\,
W^\mu{}_{\nu\rho\sigma,\gamma_1\gamma_2}\;.
\nonumber
\end{align}

\section{Conformal Gravity with Einstein sector in 4D and 6D}
\label{sec:four and six}

In four dimensions there is a single conformal invariant with mass 
dimension four, which is the square of the Weyl tensor. 
The equations of motion of the corresponding action set to zero 
the Bach tensor defined in dimension $D>3$ by
\begin{equation}
B_{\mu\nu} = \frac{1}{3-D}\,
\nabla^{\beta}\nabla^{\alpha}W_{\alpha\mu\nu\beta} 
- K^{\alpha\beta}\,W_{\alpha\mu\nu\beta} \equiv 
\frac{1}{3-D}\,{\cal D}^\beta{\cal D}^\alpha W_{\alpha\mu\nu\beta}\;.
\end{equation}
On Einstein manifolds, the Ricci tensor is proportional to the metric, 
and so is the Schouten tensor, showing that the Bach tensor vanishes 
on Einstein manifolds.
By using the differential Bianchi identity for the Riemann tensor, 
it is also easy to see that the Bach tensor is symmetric.
It is evidently traceless, 
which can be viewed as the Noether identity for the Weyl-invariance 
of $4D$ conformal gravity.  
Therefore, all the solutions of four-dimensional Einstein gravity 
(with or without cosmological constant) are solutions 
of four-dimensional conformal gravity. As we mentioned in the introduction, 
the converse is not true, and Maldacena \cite{Maldacena:2011mk} showed what 
boundary conditions to impose on the metric in asymptotically 
anti-de Sitter (AdS) manifolds in order to kill the unwanted degrees of freedom, 
leaving only those of Einstein gravity. 
Thus, upon using such boundary conditions, four-dimensional conformal 
gravity is equivalent to ordinary four-dimensional Einstein 
gravity in asymptotically AdS spacetime.

In \cite{Lu:2013hx} the most general six-dimensional conformal gravity theory 
was found, such that all Einstein manifolds are solutions to the equations of 
motion. This is less trivial than the four-dimensional case, since in six 
dimensions there are three independent Weyl invariant scalar densities, with mass 
dimension six, built with the Weyl tensor and its covariant derivatives, so that 
the six-dimensional conformal gravity depends on two free parameters, up to an 
overall constant. In the following, we review this result.

Using the Weyl-covariant tensor calculus introduced in \cite{Boulanger:2004eh}, 
it is not a difficult task to build a basis of Weyl-invariant scalar densities 
in $6D$. One finds the following three Weyl-invariant scalar densities:
\begin{align}
{\cal I}_1 &= \sqrt{|g|}\left( W_\alpha{}^{\rho\sigma}{}_\beta\,W_{\mu\rho\sigma\nu}\,
W^{\mu\alpha\beta\nu}\right)\;,
\label{I1}\\
{\cal I}_2 &= \sqrt{|g|}\left(W_{\alpha\beta}{}^{\mu\nu}\,
W_{\mu\nu\rho\sigma}\,W^{\rho\sigma\alpha\beta}\right)\;,
\label{I2}\\
\widetilde{\cal I}_3 & = \sqrt{|g|}\left(\tfrac{1}{2}\,
{\cal D}_\alpha W_{\mu\nu\rho\sigma}\,{\cal D}^\alpha W^{\mu\nu\rho\sigma}
+ \tfrac{8}{9}\,
{\cal D}_{\alpha}W^{\alpha\beta\gamma\delta}\,
{\cal D}^{\mu}W_{\mu\beta\gamma\delta} 
+ W^{\mu\nu\rho\sigma}\,
{\cal D}_{\alpha}{\cal D}^{\alpha}
W_{\mu\nu\rho\sigma}\right)\;.
\label{I3tilde}
\end{align} 
Therefore, up to boundary terms, the most general action for six-dimensional 
conformal gravity can be written as
\begin{equation}
S_6[g_{\mu\nu}] = \int\diff^6x\, (w_1\,{\cal I}_1 + w_2\,{\cal I}_2 
+ w_3\,\widetilde{\cal I}_3)\;,
\end{equation}
where the coefficients $w_i\,$, $i=1,2,3$, are arbitrary 
(non-simultaneously vanishing) real constants. 

We now compute the variation of the action $S_6[g_{\mu\nu}]\,$, 
in order to determine for which choice of coefficients $\{w_i\}$ an 
Einstein metric can be solution to the Euler-Lagrange equations of motion. 
Discarding terms that identically vanish on an Einstein manifold, we find
\begin{align}
\tfrac{1}{\sqrt{|g|}}\,\delta S_6 =& ~\Big[-\tfrac{2}{15}\,
(69\,w_1 - 144\,w_2 + 206\,w_3)\, 
  K_\mu{}^\mu\,W_\alpha{}^{\gamma\delta\epsilon}\,
W_{\beta\gamma\delta\epsilon} \,
\nonumber\\ 
& + \tfrac{3}{5}\,(19\,w_1 - 44\,w_2 + 56\,w_3)\,
W_\alpha{}^{\gamma\delta\epsilon}\,W_{\beta}{}^\nu{}_\delta{}^\mu\,W_{\gamma\mu\epsilon\nu} \,+\nonumber\\
& + \tfrac{3}{10}\,(19\, w_1 - 44\, w_2 + 56\, w_3) \,
W_\alpha{}^{\gamma\delta\epsilon}\,W_{\beta\gamma}{}^{\mu\nu}\,W_{\delta\mu\epsilon\nu} \,+\nonumber\\
& + \tfrac{1}{10}\,(-39\, w_1 + 84\, w_2 - 116\, w_3)\,K_\mu{}^\mu\,
{\cal D}^\nu W_\alpha{}^{\gamma\delta\epsilon}\,
{\cal D}_\gamma W_{\beta\nu\delta\epsilon} \,+\nonumber\\
& + \tfrac{1}{20}\,(-9\, w_1 + 24\, w_2 - 26\, w_3)\,K_\mu{}^\mu\,
{\cal D}_\beta W_{\gamma\delta\epsilon\nu}\,
{\cal D}_\alpha W^{\gamma\delta\epsilon\nu}\Big]\,(\delta g^{\alpha\beta} -\tfrac{1}{6}\,g_{\mu\nu}\,\delta g^{\mu\nu}\,g^{\alpha\beta}) \;.
\end{align}
One of the main tasks leading to the above expression was to write it 
in a basis of linearly independent structures, so that the expression vanishes 
if and only if its coefficients vanish. This happens if and only if
\begin{equation}
w_2 = \tfrac{1}{20}\,w_1, \quad
w_3 = -\tfrac{3}{10}\,w_1\;,
\end{equation}
so that, up to an overall constant -- choose $w_1=\tfrac{20}{3}$ -- 
the action $\tilde{S}_6[g_{\mu\nu}]$ for the six-dimensional conformal 
gravity theory with an Einstein sector is unique, equal to
\begin{equation}\label{CriticalAction6d}
\tilde{S}_6[g_{\mu\nu}] = \int\text{d}^6x\,\,(\tfrac{20}{3}\,{\cal I}_1 
+\tfrac{1}{3}\,{\cal I}_2 - 2\,\widetilde{\cal I}_3)\;.
\end{equation}
This result is consistent with the combination 
$4\,{\cal I}_1 + {\cal I}_2 - \frac{1}{3}\,{\cal I}_3$ found \cite{Lu:2013hx}, 
where $I_3 = \frac{{\cal I}_3}{\sqrt{|g|}}$ 
is the last invariant in Eq. (1.1) of \cite{Lu:2013hx}, 
since $\widetilde{I}_3 = \frac{\widetilde{\cal I}_3}{\sqrt{|g|}}$
with $\widetilde{\cal I}_3$ given in \eqref{I3tilde} can also be written in 
the following way:
\begin{align}
\widetilde{I}_3 \,=\;& \nabla^\alpha\,(\tfrac{1}{2}\,W^{\beta\gamma\delta\epsilon}\,W_{\beta\gamma\delta\epsilon,\alpha}-\tfrac{8}{9}\,W_\alpha{}^{\beta\gamma\delta}\,W_\beta{}^\epsilon{}_{\gamma\delta,\epsilon}) + \tfrac{4}{3}\,I_1 - \tfrac{1}{3}\,I_2 \,+\nonumber\\
& + \tfrac{1}{6}\,W^{\alpha\beta\gamma\delta}\,W_{\alpha\beta\gamma\delta,\epsilon}{}^{\epsilon} +\tfrac{8}{3}\,W_\alpha{}^{\gamma\delta\epsilon}\,W_{\beta\gamma\delta\epsilon}\,K^{\alpha\beta}- W_{\beta\gamma\delta\epsilon}\,W^{\beta\gamma\delta\epsilon}
\,K_\alpha{}^\alpha\;,
\end{align}
where the second line is equal to the bulk terms in $\frac{1}{6}\,I_3$
of \cite{Lu:2013hx}, so that we have
$\widetilde{\cal I}_3 = \frac{4}{3}\,{\cal I}_1 - \frac{1}{3}\,{\cal I}_2 +
\frac{1}{6}\,{\cal I}_3\, + \partial_\mu {\cal V}^{\mu}\,$,
and from it follows the equality of Lagrangian densities, up to total derivatives: 
\begin{equation}
\tfrac{20}{3}\,{\cal I}_1 + \tfrac{1}{3}\,{\cal I}_2 -
2\,\widetilde{\cal I}_3\, =
4\,{\cal I}_1 + \,{\cal I}_2 - \tfrac{1}{3}\,{\cal I}_3\, 
-2\, \partial_\mu {\cal V}^\mu\;.
\end{equation}

Finally, for completeness we note that the conformal invariant 
given in Prop. 3.4 of \cite{Graham1985} is given by
\begin{align}
\mathcal{I}_3^{\text{(FG)}} = \sqrt{|g|}\,\Big(&  
16\,C_{\alpha\beta\gamma}\,C^{\alpha\beta\gamma} + 
16\,W_\mu{}^{\alpha\beta\gamma}\,W_{\nu\alpha\beta\gamma}\,K^{\mu\nu} 
+\nabla_\varepsilon\,W_{\alpha\beta\gamma\delta}\,\nabla^\varepsilon\,
W^{\alpha\beta\gamma\delta}  
\nonumber \\
& + 16\,W_{\alpha\beta\gamma\delta}\,\nabla^\beta\,C^{\alpha\gamma\delta}
\Big)\;.
\end{align}
One can explicitly verify that the relation with $\tilde{\mathcal{I}}_3$ is 
\begin{equation}
\mathcal{I}_3^{(\text{FG})} = 2\,(\tilde{\mathcal{I}}_3 - 4\,\mathcal{I}_1 + \mathcal{I}_2).
\end{equation}

\section{Relation with Branson's $Q$-curvature}
\label{sec:Q-curvature}

The notion of $Q$-curvature was introduced by Branson when studying 
the regularisation of the functional determinant of elliptic operators 
\cite{branson1993functional}. It emerges in many other mathematical 
contexts~\cite{branson2005q} and, in particular, 
plays an important r\^ole in conformal geometry \cite{chang2008q}, 
see also the book \cite{juhl2009} and refs. therein.

One may introduce the $Q$-curvature by studying how to complete 
the powers of the Laplacian -- in this section we assume the manifold 
to be Riemannian, but the signature will be irrelevant to the discussion -- 
to obtain a conformally covariant operator.
An operator $\mathcal{O}$ is said to be conformally covariant if it 
transforms under infinitesimal Weyl transformation in the following way:
\begin{equation}\label{ConformalOperator}
\delta_\sigma\,\mathcal{O}\,\vphi = \beta\,\sigma\,\mathcal{O}\,\vphi\;, \quad 
\text{if}\;\;\delta_\sigma\,\vphi = \alpha\,\sigma\,\vphi\;, \quad 
\text{for some constants}\;\alpha\,, \beta\;.
\end{equation}
In $D$ dimensions the transformation of the Laplacian 
$\bigtriangleup = g^{\mu\nu}\,\nabla_\mu\nabla_\nu$ is
\begin{equation}
\delta_\sigma\,{\bigtriangleup\vphi} = 
-(2-\alpha)\,\sigma\,{\bigtriangleup\vphi} + \alpha\,{\bigtriangleup\sigma}\,
\vphi + (2\,\alpha - 2 + D)\,\,\nabla_\mu\,\sigma\,\nabla^\mu\,\vphi\;,
\end{equation}
if $\vphi$ transforms as in \eqref{ConformalOperator}. 
There is no choice of $\alpha$ to make it conformally covariant. 
But one can notice that the Laplacian of the Weyl parameter is included 
in the transformation of the trace of the Schouten tensor:
\begin{equation}\label{TransformationK}
\delta_\sigma\,K_\mu{}^\mu = -{\bigtriangleup\sigma} - 2\,\sigma\,K_\mu{}^\mu \;,
\end{equation}
so that, for some constant $\beta$, 
\begin{align}
\delta_\sigma\,(\bigtriangleup + \beta\,K_\mu{}^\mu )\,\vphi =&\, -(2-\alpha)\,\sigma\,(\bigtriangleup + \beta\,K_\mu{}^\mu )\,\vphi \,+\nonumber \\
& +(\alpha-\beta)\,{\bigtriangleup\sigma}\,\vphi  + (2\,\alpha -2 + D)\,\nabla_\mu\,\sigma\,\nabla^\mu\,\vphi\;.
\end{align}
Thus, is it sufficient to choose $\alpha = \beta = -\frac{D-2}{2}$ 
to get a conformally covariant operator. The resulting operator in $D$ 
dimensions, usually called \emph{Yamabe operator} \cite{yamabe1960deformation}, 
reads
\begin{equation}
Y_D = \bigtriangleup + \tfrac{D-2}{2}\,K_\mu{}^\mu = 
 \bigtriangleup + \tfrac{D-2}{4(D-1)}\,R\;, 
\end{equation}
whose transformation is
\begin{equation}
\delta_\sigma\,Y_D\,\vphi = -\tfrac{D+2}{2}\,\sigma\,Y_D\,\vphi\;,
\quad \text{if}\;\;\delta_\sigma\,\vphi = -\tfrac{D-2}{2}\,\sigma\,\vphi\;.
\end{equation}
Notice that in the critical dimension $D=2$, the Laplacian is automatically 
conformally covariant, and the integral of the density $\sqrt{|g|}\,K$ 
(or equivalently, the Einstein-Hilbert action) is conformally invariant, 
since the Laplacian contribution ${\bigtriangleup\sigma}$ in the 
conformal transformation of \textbf{$K_\mu{}^\mu $}
contributes through a total derivative.

Consider now the more ambitious task of conformally completing the 
square of the Laplacian. 
The result in four dimensions was found by Fradkin and Tseytlin in 
\cite{Fradkin:1982xc}, and also by Riegert \cite{Riegert:1984kt}; 
in arbitrary dimensions, it was found by Paneitz in 
\cite{Paneitz:2008afy}. 
By dimensional analysis, one can start from the 
following ansatz:\footnote{Recall that $\nabla^\mu\,K_{\mu\nu} = \nabla_\nu\,K\,$, 
as a consequence of the differential Bianchi identity.}
\begin{align}\label{AnsatzPaneitz}
P_D\,\vphi =\;& {\bigtriangleup^2\vphi} 
+ \beta_1\,\nabla^\nu\,K_\mu{}^\mu \,\nabla_\nu\,\vphi 
+ \beta_2\,{\bigtriangleup K_\mu{}^\mu }\,\vphi 
+ \beta_3\,K_{\mu\nu}\,K^{\mu\nu}\,\vphi \,+\nonumber\\
& + \beta_4\,K_\mu{}^\mu\,K_\nu{}^\nu \,\vphi 
+ \gamma_1\,K^{\mu\nu}\,\nabla_\mu\,\nabla_\nu\,\vphi 
+ \gamma_2\,K_\mu{}^\mu \,{\bigtriangleup\vphi}\;.
\end{align}
By explicit evaluation, one obtains
\begin{align}
\delta_\sigma\,P_D\,\vphi + (4-\alpha)\,\sigma\,P_D\,\vphi 
\,=\;& 
(-2\,(\beta_1+\beta_2) + \alpha\,\gamma_2)\,K_\mu{}^\mu\,
{\bigtriangleup\sigma}\;\vphi 
\nonumber\\
& + (\alpha\,\beta_2 + (D-6)\,\beta_2)\,\nabla_\mu 
K_\nu{}^\nu\,\nabla^\mu \sigma\;\vphi 
\nonumber\\
& + (6-\beta_1 + \gamma_1 + 2\,\alpha\,(\gamma_2 -2) + D\,(\gamma_2 -1) - 2\,\gamma_2)\,
K_\mu{}^\mu\,\nabla_\nu \sigma\,\nabla^\nu\;\vphi 
\nonumber \\
& + (-2\,\beta_3 + \alpha\,\gamma_1)\,
K^{\mu\nu}\,\nabla_\mu \nabla_\nu \sigma\;\vphi \nonumber\\
& + (2\,(\alpha - 1) - \gamma_2)\,{\bigtriangleup\sigma}\,{\bigtriangleup\vphi} 
 + 2\,(D-4+2,\alpha)\,\nabla^\mu \sigma\,{\bigtriangleup\nabla_\mu \vphi} 
\nonumber\\
& + (D-2+4\,\alpha-\beta_1)\,\nabla^\mu\vphi\,{\bigtriangleup\nabla_\mu \sigma} 
 + (\alpha-\beta_2)\,\vphi\,{\bigtriangleup^2\sigma} \,
\nonumber \\
& + ((D-2)(D-6+4\,\alpha - \beta_1) + 2\,
(\alpha-1)\,\gamma_1)\,K^{\mu\nu}\,\nabla_\mu \sigma\,\nabla_\nu \sigma 
\nonumber\\
& + (2\,(D-2) + 4\,\alpha - \gamma_1)
\,\nabla_\mu \nabla_\nu \sigma\,\nabla^\mu \nabla^\nu \vphi\;.
\end{align}
The right-hand side vanishes if and only if
\begin{equation}
\alpha = \beta_2 = -\tfrac{D-4}{2}, \;\;
\beta_1 = 6-D, \;\;
\beta_3 = 4-D, \;\;
\beta_4 = \tfrac{D(D-4)}{4}, \;\;
\gamma_1 = 4, \;\;
\gamma_2 = 2-D.
\end{equation}
Replacing these values for the constants in the ansatz \eqref{AnsatzPaneitz}, 
and manipulating a little bit, one finds the \emph{Paneitz operator}
\begin{align}
P_D =\;& \nabla_\mu\,(\nabla^\mu\,\nabla^\nu + 4\,K^ {\mu\nu} 
- 4\,(D-2)\,g^{\mu\nu}\,K_\rho{}^\rho)\,\nabla_\nu\, 
\nonumber \\
& + \tfrac{D-4}{2}\,(-2\,K_{\mu\nu}\,K^{\mu\nu} + \tfrac{D}{2}\,K_\mu{}^\mu\,K_\nu{}^\nu - {\bigtriangleup K_\mu{}^\mu })\,\;.
\end{align}
We recognise the same structure as in the Yamabe operator $Y_D\,$: 
on the first line there is the Laplacian squared (improved in such 
a way as to take the form 
$\nabla_\mu\,\mathcal{S}^{\mu\nu}\,\nabla_\mu$, where $\mathcal{S}^{\mu\nu}$ 
is rank-two symmetric tensor operator), while the second line, 
which vanishes in four dimensions, gives a purely multiplicative 
(i.e., non-differential) operator. 
Let us denote it by ${\cal Q}_{4,D}$, where $4$ is the order 
of $\bigtriangleup^2\,$:
\begin{equation}\
{\cal Q}_{4,D} := \sqrt{|g|}\,(-2\,K_{\mu\nu}\,K^{\mu\nu} + \tfrac{D}{2}\,K_\mu{}^\mu\,K_\nu{}^\nu - {\bigtriangleup K_\mu{}^\mu })\;,
\end{equation}
the density factor $\sqrt{|g|}$ being included for future convenience. 
In analogy with the Yamabe operator case, ${\cal Q}_{4,D}$ is expected to be 
conformally invariant, when integrated on a closed manifold of 
dimension $D=4\,$: 
\begin{equation}
\delta_\sigma\int \diff^4 x\,{\cal Q}_4 = \delta_\sigma\int\diff^4 x\,\sqrt{|g|}
(-2\,K_{\mu\nu}\,K^{\mu\nu} + 2\,K_\mu{}^\mu\,K_\nu{}^\nu  
- {\bigtriangleup K_\mu{}^\mu }) = 0\;,
\end{equation}
where ${\cal Q}_4 := {\cal Q}_{4,4}$. 
A simple way to see this is to notice that 
\begin{align}
\tfrac{1}{\sqrt{-g}}\,{\cal Q}_4 &= -2\,K_{\mu\nu}\,K^{\mu\nu} 
+ 2\,K_\mu{}^\mu\,K_\nu{}^\nu  - {\bigtriangleup K_\mu{}^\mu }   
\label{Q4}\\
& = -\tfrac{1}{16}\,(32\,(K_{\mu\nu}\,K^{\mu\nu}-K_\mu{}^\mu\,K_\nu{}^\nu ) -4\,W_{\mu\nu\rho\sigma}\,W^{\mu\nu\rho\sigma}) -\tfrac{1}{4}\,W_{\mu\nu\rho\sigma}\,W^{\mu\nu\rho\sigma}- {\bigtriangleup K_\mu{}^\mu} 
\nonumber \\
&= -\tfrac{1}{16}\,\varepsilon^{\mu\nu\rho\sigma}\,\varepsilon_{\alpha\beta\gamma\delta}\,R_{\mu\nu}{}^{\alpha\beta}\,R_{\rho\sigma}{}^{\gamma\delta} -\tfrac{1}{4}\,W_{\mu\nu\rho\sigma}\,W^{\mu\nu\rho\sigma}
- {\bigtriangleup K_\mu{}^\mu }\;,\label{DecompositionQ4}
\end{align}
where the first term on the last line 
is the four-dimensional Euler invariant 
(\emph{Gauss-Bonnet invariant}), which is topological. 
Therefore, when integrated, only the second term could contribute 
to the Weyl transformation, but it is manifestly 
conformally invariant in 4$D$ when multiplied by $\sqrt{|g|}$
to make it a scalar density.

This story can be generalised in the following way. 
One considers the $m$th power $\bigtriangleup^m$ of the Laplacian; 
its conformal completion $P_{2m,D}$ in $D$ dimensions was discussed by 
Graham, Jenne, Mason, and Sparling, in \cite{graham1992conformally}. 
As argued by Branson, it takes the form
\begin{equation}
P_{2m,D} = \nabla_\mu\,\mathcal{S}_D^{\mu\nu}\,\nabla_\nu 
+ \tfrac{D-2m}{2}\,\tfrac{1}{\sqrt{|g|}}\,{\cal Q}_{2m,D}\;,
\end{equation}
where $\mathcal{S}_D^{\mu\nu}$ is a rank-two symmetric tensor operator, 
such that 
$\nabla_\mu\,\mathcal{S}_D^{\mu\nu}\,\nabla_\nu  = \bigtriangleup^m + \dots$, 
where the ellipsis stands for lower derivative terms, 
and ${\cal Q}_{2m,D}$ is a purely multiplicative (non-differential) 
operator defined by the explicit expression of $P_{2m,D}\,$. 
The conformal transformation of $P_{2m,D}$ is required to be
\begin{equation}
\delta_\sigma\,P_{2m,D}\,\vphi = -\tfrac{D+2m}{2}\,\sigma\,P_{2m,D}\,\vphi\;, \quad\text{if}\;\;
\delta_\sigma\,\vphi = -\tfrac{D-2m}{2}\,\sigma\,\vphi\;.
\end{equation}
In $D=2\,m$, ${\cal Q}_{2m} := {\cal Q}_{2m,2m}$ is the 
$Q$\emph{-curvature} in $2m$ dimensions. 
Its integral is conformally invariant:
\begin{equation}
\delta_\sigma\int\diff^{2m}\,x\,{\cal Q}_{2m} = 0\;.
\end{equation}
If $m=1$, $P_{2,D} = Y_D$ is the Yamabe operator, 
and ${\cal Q}_{2,D} = -\sqrt{|g|}\,K_\mu{}^\mu\,$, 
so that the two-dimensional $Q$-curvature is
\begin{equation}
{\cal Q}_2 = -\sqrt{|g|}\,K_\mu{}^\mu = -\tfrac{1}{2}\,\sqrt{|g|}\,R\;.
\end{equation} 
If $m=2$, $P_{4,D} = P_D$ is the Paneitz operator, 
and the four dimensional $Q$-curvatures is given by \eqref{Q4}. 
The property \eqref{DecompositionQ4} for the decomposition of a 
global conformal invariant generalises to arbitrary even 
dimensions\footnote{In \cite{Boulanger:2018rxo}, 
a confusion in the motivation behind the works leading to 
\cite{alexakis2012decomposition} is explained. The conjecture 
made by Deser and Schwimmer \cite{Deser:1993yx}, 
taken as a motivation in \cite{alexakis2012decomposition}, 
does \emph{not} concern global conformal invariants. 
Instead, it concerns the general structure of conformal (or Weyl) anomalies
in quantum field theory, 
a different notion as compared to global conformal invariants. 
The conjecture \cite{Deser:1993yx} of Deser and Schwimmer for the 
classification of conformal anomalies was solved in 
\cite{Boulanger:2007ab,Boulanger:2007st} by using cohomological 
techniques. In particular, it was proven that conformal anomalies 
are trivial in odd spacetime dimensions. 
The same cohomological tools were used in \cite{Boulanger:2018rxo} 
to provide an alternative derivation and completion of the main 
result of \cite{alexakis2012decomposition} 
concerning the general structure of global conformal invariants
in arbitrary dimension.
In particular, in \cite{Boulanger:2018rxo} were found the global 
conformal invariants in dimensions $4m-1\,$, $m\in \mathbb{N}^+\,$.} 
\cite{alexakis2012decomposition,Boulanger:2018rxo}: 
\begin{equation}\label{FormulaQcurvature}
{\cal Q}_{2m} = \alpha_D\;{\cal E}_{2m}(R) +  \mathcal{I} + \partial_\mu\,\mathcal{V}^\mu,
\end{equation}
where $\alpha_D$ is a constant, 
${\cal E}_{2m}(R)$ is the Euler density in dimension $D=2m$,
\begin{equation}\label{EulerInvariant}
{\cal E}_{2m}(R) = 
\sqrt{|g|}\,
\varepsilon^{\mu_1\dots\mu_{2m}}\,
\varepsilon_{\alpha_1\dots\alpha_{2m}}\,
R_{\mu_1\mu_2}{}^{\alpha_1\alpha_2}\dots 
R_{\mu_{2m-1}\mu_{2m}}{}^{\alpha_{2m-1}\alpha_{2m}}\;,
\end{equation} 
and where the second term $\mathcal{I}$ is a \emph{local} (i.e. pointwise) 
conformally invariant density. 
The general decomposition \eqref{FormulaQcurvature} implies that the 
functional derivative 
$\frac{1}{\sqrt{|g|}}\frac{\delta S_{2m}}{\delta g^{\mu\nu}}$
of the functional 
$S_{2m}[g]=\int\diff^{2m} x\,{\cal Q}_{2m}$ furnishes a 
divergenceless, traceless, rank-two symmetric, conformally 
covariant tensor of weight $2-2m\,$. 
In two dimensions it is obviously proportional to the 
Einstein tensor $G_{\mu\nu}=R_{\mu\nu}-\frac{1}{2}g_{\mu\nu}R\,$. 
In four dimensions, it gives the Bach tensor $B_{\mu\nu}$, 
the left-hand side of the 
equations of motion of four-dimensional conformal gravity:
\begin{align}
\delta\int \diff^4 x\,{\cal Q}_4 &= \delta \left( -\frac{1}{4}\int 
\diff^4 x\sqrt{|g|}\,W_{\mu\nu\rho\sigma}\,W^{\mu\nu\rho\sigma} \right) 
\nonumber\\
& = -\int\diff^4 x\,\sqrt{|g|}\,(\nabla^\alpha\,C_{\mu\nu\alpha} 
+ W_{\mu\alpha\nu\beta}\,K^{\alpha\beta})\,\delta g^{\mu\nu} 
\nonumber \\
& = -\int\diff^4 x\,\sqrt{|g|}\,B_{\mu\nu}\,\delta g^{\mu\nu}.
\end{align}
In the general case, as proved in \cite{graham2005ambient}, one gets a 
higher-dimensional generalisation of the Bach tensor, called the
\emph{Fefferman-Graham obstruction tensor} ${O}^{(2m)}_{\mu\nu}$, 
introduced in the context of the ambient metric construction of  
\cite{Graham1985} -- see also \cite{chang2008q} for a concise review 
of the $Q$-curvature and its definition in terms of the ambient metric
in dimension $2m+2$.
Explicitly, 
\begin{equation}
\delta\int \diff^{2m} x\;{\cal Q}_{2m} = 
-\int\diff^{2m} x\,\sqrt{|g|}\,{O}^{(2m)}_{\mu\nu}\,\delta g^{\mu\nu}\;,
\end{equation}
where ${O}^{(2)}_{\mu\nu} = -\tfrac{1}{2}\,G_{\mu\nu}$, 
and ${O}^{(4)}_{\mu\nu} = B_{\mu\nu}\,$.

The Fefferman-Graham obstruction tensor ${O}^{(2m)}_{\mu\nu}$ 
is not only divergenceless, traceless and symmetric, 
but it also enjoys the property that it identically vanishes for 
metrics that are conformally Einstein, see e.g. \cite{graham2005ambient},  
also Chapt. 7 of \cite{fefferman2012ambient}, and references therein. 
Thus, the local conformal invariant $\mathcal{I}$ in the general 
decomposition \eqref{FormulaQcurvature} of the $Q$-curvature 
is a combination of the possible $2m$-dimensional local 
(pointwise) conformal invariants, such that the variational 
principle based on ${\cal Q}_{2m}$ always admits an Einstein 
sector upon extremization.
Since the number of local conformal invariants quickly growths 
with the dimension, in general there might be several linear 
combinations of the local conformal invariants that lead to 
symmetric, divergenceless and traceless tensors vanishing on 
Einstein metrics, and the integrated $Q$-curvature could be only one 
among many global conformal invariants that give rise to such 
symmetric tensors, upon variational derivative with respect 
to the (inverse) metric.
Equivalently, in an arbitrary space of even dimension $D=2m$, 
there could be several tensors that share the properties of 
the Fefferman-Graham obstruction tensor. We will return to this discussion 
in Section \ref{sec:Discu} and proceed now with a detailed review of the 
six-dimensional case.

The six-dimensional $Q$-curvature can be computed following the above 
construction. The result is \cite{Gover:2002ay} 
(see also \cite{branson2005q}):
\begin{align}
{\cal Q}_6 = \sqrt{|g|}\,\Big(& 8\,\nabla^\mu K^{\nu\rho}\,
\nabla_\mu K_{\nu\rho} 
+ 16\,K_{\mu\nu}\,\bigtriangleup K^{\mu\nu} 
- 32\,K_{\mu\nu}\,K^\mu{}_\rho\,K^{\nu\rho} \,
\nonumber\\
& - 16\,K^{\mu\nu}\,K_{\mu\nu}\,K_\rho{}^\rho + 8\,K_\mu{}^\mu\,K_\nu{}^\nu\,K_\rho{}^\rho 
- 8\,K_\mu{}^\mu\,\bigtriangleup K_\nu{}^\nu \,\nonumber\\
& + \bigtriangleup^2\,K_\mu{}^\mu + 16\,W^{\mu\rho\nu\sigma}\,K_{\mu\nu}\,K_{\rho\sigma}\Big)\;.
\label{Q6}
\end{align}
The explicit expression for its variation, given by the six-dimensional 
Fefferman-Graham obstruction tensor, is explicitly computed in  
\cite{graham2005ambient}:
\begin{align}
O^{(6)}_{\mu\nu} = -\tfrac{1}{2}\,(& \bigtriangleup B_{\mu\nu} 
- 2\,W_{\rho\mu\nu\sigma}\,B^{\rho\sigma} - 4\,B_{\mu\nu}\,K_\rho{}^\rho \,
+ 8\,\nabla_\sigma\,C_{\mu\nu\rho}\,K^{\rho\sigma} 
\nonumber\\
&+ 8\,\nabla_\sigma\,C_{\nu\mu\rho}\,K^{\rho\sigma} 
- 4\,C^\rho{}_\mu{}^\sigma\,C_{\sigma\nu\rho} 
+ 2\,C_\mu{}^{\rho\sigma}\,C_{\nu\rho\sigma} 
+ 4\,\nabla_\sigma\,K_\rho{}^\rho\,W_{\mu\nu}{}^\sigma 
\nonumber\\
& + 4\,\nabla_\sigma\,K_\rho{}^\rho\,W_{\nu\mu}{}^\sigma - 4\,W_{\rho\mu\nu\sigma}\,K_\tau{}^\rho\,K^{\sigma\tau})\;,
\end{align}
which is divergenceless, traceless, symmetric, and identically 
vanishing on Einstein manifolds, as it should be. 
Consistently with the general structure \eqref{FormulaQcurvature}, 
one can show (see also \cite{Anastasiou:2018mfk, Lu:2019urr,Aros:2019tjw}) 
that ${\cal Q}_6$ in \eqref{Q6} can equivalently be written as
\begin{align}
{\cal Q}_6 =&\, - \tfrac{10}{3}\,\mathcal{I}_1 -\tfrac{1}{6}\,\mathcal{I}_2 
+ \widetilde{\mathcal{I}}_3 -\tfrac{1}{48}\,{\cal E}_6(R) 
\nonumber\\
& + \sqrt{|g|}\;\nabla^\mu\,\Big(5\,W_{\mu\alpha\beta\gamma}\,C^{\alpha\beta\gamma} 
+ 8\,K_\mu{}^\alpha\,\nabla_\alpha\,K_\beta{}^\beta 
+ W^{\alpha\beta\gamma\delta}\,\nabla_\beta\,W_{\mu\alpha\gamma\delta} 
\nonumber\\
& - 8\,K^{\alpha\beta}\,\nabla_\beta\,K_{\mu\alpha} 
+ 16\,K^{\alpha\beta}\,\nabla_\mu\,K_{\alpha\beta} 
- 8\,K_\nu{}^\nu\,\nabla_\mu\,K_\rho{}^\rho 
+ \nabla_\mu\bigtriangleup K_\nu{}^\nu
\Big)\;,
\end{align}
where the six-dimensional Euler density is, according to \eqref{EulerInvariant},
\begin{equation}
{\cal E}_6(R) 
= \sqrt{|g|}\,
\varepsilon^{\mu\nu\rho\sigma\kappa\lambda}\,\varepsilon_{\alpha\beta\gamma\delta\varepsilon\zeta}\,R_{\mu\nu}{}^{\alpha\beta}\,R_{\rho\sigma}{}^{\gamma\delta}\,R_{\kappa\lambda}{}^{\varepsilon\zeta}\;.
\end{equation}
Since the Euler invariant is topological and the last two lines 
contribute to a total derivative, the integral of the $Q$-curvature 
on a closed manifold is proportional 
to the conformal action $\tilde{S}_6[g_{\mu\nu}]$ \eqref{CriticalAction6d}
with Einstein sector, up to an additive constant arising from 
the integral of the Euler density:
\begin{align}
\int_{M_6} \diff^6 x\;{\cal Q}_6 &= 64\pi^3\,\chi(M_6)
+\int_{M_6} \diff^6 x\,\sqrt{|g|}\,(-\tfrac{10}{3}\,
\mathcal{I}_1 -\tfrac{1}{6}\,\mathcal{I}_2 + \widetilde{\mathcal{I}}_3)\;,
\end{align}
where one uses
\begin{equation}
\int_{M_{2m}}\diff^{2m} x\,\mathcal{E}_{2m}(R) = (-1)^m\,(4\,\pi)^m\,m!\,2^m\,\chi(M_{2m}),
\end{equation}
and one recognises the precise combination which defines the 
L{\"u}-Pang-Pope 
six-dimensional conformal gravity action in Eq. \eqref{CriticalAction6d}.
That is, up to boundary terms, one has
\begin{equation}
\tilde{S}_6[g_{\mu\nu}] = 128\pi^3\,\chi(M_6) \,- \,2\int_{M_6} 
\diff^6 x\;{\cal Q}_6\;.
\end{equation}

\section{8D Conformal Gravity with Einstein Sector}
\label{sec:8D}

In this section, we build the most general conformal gravity action 
in 8$D$ that admits an Einstein sector.
We will see that, although the number of local conformal invariants 
increases dramatically compared to the 4$D$ and 6$D$ cases, there 
still is only one linear combination of them that ensures that 
the theory admits an Einstein sector, and we will see that this 
reproduces the $Q$-curvature in eight dimensions.

There are seven possible parity-even scalars that are quartic in the 
undifferentiated Weyl tensor. 
One can choose the following basis \cite{Fulling:1992vm}:
\begin{align}
I_6 &= W_{\alpha\beta}{}^{\nu\sigma}\,W^{\alpha\beta\gamma\delta}\,
W_{\gamma\nu}{}^{\rho\mu}\,W_{\delta\sigma\rho\mu}\;,\\
I_7 &= W_\alpha{}^\nu{}_\gamma{}^\sigma\,W^{\alpha\beta\gamma\delta}\,W_\beta{}^\rho{}_\delta{}^\mu\,W_{\nu\rho\sigma\mu}\;,\\
I_8 &= W_{\alpha\beta}{}^{\nu\sigma}\,W^{\alpha\beta\gamma\delta}\,W_{\gamma\delta}{}^{\rho\mu}\,W_{\nu\rho\sigma\mu}\;,\\
I_9 &= W_{\alpha\beta\gamma}{}^\nu\,W^{\alpha\beta\gamma\delta}\,W_\delta{}^{\sigma\rho\mu}\,W_{\nu\rho\sigma\mu}
\;,\\
I_{10} &= W_{\alpha\beta\gamma\delta}\,W^{\alpha\beta\gamma\delta}\,W_{\nu\rho\sigma\mu}\,W^{\nu\sigma\rho\mu}\;,\\
I_{11} &= W_\alpha{}^\nu{}_\gamma{}^\sigma\,W^{\alpha\beta\gamma\delta}\,W_\beta{}^\rho{}_\sigma{}^\mu\,W_{\delta\mu\nu\rho}\;,\\
I_{12} &= W_{\alpha\gamma}{}^{\nu\sigma}\,W^{\alpha\beta\gamma\delta}\,W_\beta{}^\rho{}_\nu{}^\mu\,W_{\delta\mu\sigma\rho}\;.
\end{align}
The corresponding densities ${\cal I}_i = \sqrt{|g|} \,I_i\,$, $i\in \{6, \ldots ,12\}\,$, 
are trivially Weyl-invariant in $8D\,$. 
Then, there are five independent non-trivial Weyl-invariant scalar densities
${\cal I}_j = \sqrt{|g|} \,I_j\,$, $j\in \{1, \ldots ,5\}\,$, 
in eight-dimensions \cite{Boulanger:2004zf}, that involve derivatives 
of the Weyl tensor.
In total, that gives twelve linearly independent, 
local (i.e. pointwise) conformal invariants in $8D\,$.
As a result of the findings in \cite{Chen:2024kuw}, we find that two of 
the five non-trivial invariants of \cite{Boulanger:2004zf}, 
namely ${\cal I}_4$ and ${\cal I}_5$, 
can be expressed in terms of the other ten, up to total derivatives.
Therefore, if one is interested in the problem of integrated densities
and consider a closed 8$D$ manifold, 
the two invariants ${\cal I}_4$ and ${\cal I}_5$ from 
the list of \cite{Boulanger:2004zf} can be omitted.
More in details, we find that the two independent, dimension-eight, 
Weyl-invariant total derivatives found in \cite{Chen:2024kuw} can be written 
in terms of the $W$-tensors as $\sqrt{|g|}\,\,\nabla_\mu\,J_{(i)}^\mu(W)\,$, 
$i=1,2\,$, where
\begin{align}
J_{(1)}^\alpha(W) =&\, -\tfrac{1}{5}\,W_{\beta\gamma\delta}{}^{\sigma}\,W^{\beta\gamma\delta\epsilon}\,
{\cal D}_\rho{\cal D}^\rho W^{\alpha}{}_{\epsilon\sigma}
+W^{\alpha\beta\gamma\delta}\,W_{\beta}{}^{\epsilon\sigma\rho}\,
{\cal D}_\rho W_{\gamma\delta\epsilon\sigma}
\nonumber \\
& -\tfrac{4}{15}\,
W^{\alpha\beta\gamma\delta}\,W_\beta{}^\epsilon{}_\gamma{}^\sigma\,
{\cal D}^{\rho}W_{\delta\epsilon\sigma\rho}\,
+\tfrac{8}{15}\,W^{\alpha\beta\gamma\delta}\,W_\beta{}^\epsilon{}_\gamma{}^\sigma\,
{\cal D}^{\rho}W_{\epsilon\sigma\delta\rho}
\nonumber \\
& +\tfrac{4}{15}\,W^{\alpha\beta\gamma\delta}\,W_\beta{}^\epsilon{}_\gamma{}^\sigma\,
{\cal D}^{\rho}W_{\delta\sigma\epsilon\rho}\;,
\\
J_{(2)}^\alpha(W) =&\; 
W^{\alpha\beta\gamma\delta}\,W_\gamma{}^{\epsilon\sigma\rho}\,
{\cal D}_{\rho}W_{\beta\epsilon\delta\sigma}-W^{\alpha\beta\gamma\delta}\,
W_{\beta}{}^{\epsilon\sigma\rho}\, {\cal D}_{\rho} W_{\gamma\delta\epsilon\sigma}\,
\nonumber\\
& -\tfrac{2}{5}\,W^{\alpha\beta\gamma\delta}\,W_\beta{}^\epsilon{}_\gamma{}^\sigma\,
{\cal D}^{\rho}W_{\epsilon\sigma\delta\rho}
-\tfrac{2}{5}\,
W^{\alpha\beta\gamma\delta}\,W_\beta{}^\epsilon{}_\gamma{}^\sigma\,
{\cal D}^{\rho}W_{\delta\sigma\epsilon\rho}\;.
\end{align}
For a different proof that the space of local conformal invariants 
of weight $-8$ that are divergences on 8-manifolds is 2-dimensional, 
see \cite{Case:2024oih}.

Then, we find the following relations:
\begin{align}
I_4 &= \tfrac{1}{40}\,I_2 - \tfrac{1}{40}\,I_3 + \tfrac{25}{3}\,I_6 + \tfrac{8}{3}\,I_7 + \tfrac{2}{3}\,I_8 - 7\,I_9 - \tfrac{8}{3}\,I_{11} - \tfrac{58}{3}\,I_{12} + \nabla_\alpha\,(-5\,J_{(1)}^\alpha + 2\,J_{(2)}^\alpha)\;,\\
I_5 &= \tfrac{1}{5}\,I_2 + \tfrac{14}{3}\,I_6 + \tfrac{4}{3}\,I_7 + \tfrac{1}{3}\,I_8 - 4\,I_9 - \tfrac{4}{3}\,I_{11} - \tfrac{32}{3}\,I_{12} - 4\,\nabla_\alpha\,J_{(1)}^\alpha\;,
\end{align}
that allow us to omit the two densities ${\cal I}_4$ and ${\cal I}_5$ from 
the expression for the Lagrangian density of conformal gravity in $8D\,$.
The remaining three non-trivial invariants of 
\cite{Boulanger:2004zf} 
are recalled here, for the sake of completeness:
\begin{align}
I_1 =\;& W_{\rho\gamma\mu\sigma}\,
W^{\rho\gamma\mu\sigma,\alpha}{}_{\alpha\beta}{}^\beta
+\tfrac{48}{25}\,W^\beta{}_{\gamma\mu\alpha,\beta}
W^{\rho\gamma\mu\alpha}{}_{\rho\nu}{}^\nu \,
\nonumber \\
& +2\,W_{\mu\beta\gamma\nu,\alpha}\,W^{\mu\beta\gamma\nu,\alpha\rho}{}_{\rho} 
+ \tfrac{42}{125}\,W_{\gamma\alpha\beta\mu}{}^{\beta\alpha}\,
W^{\gamma\nu\rho\mu}{}_{\rho\nu} \,
\nonumber \\
& +\tfrac{9}{10}\,W_{\alpha\mu\nu\beta,\gamma}{}^\gamma\,
W^{\alpha\mu\nu\beta,\rho}{}_\rho
+\tfrac{3}{5}\,W_{\nu\gamma\mu\rho,\beta\alpha}\,W^{\nu\gamma\mu\rho,\beta\alpha} \,
\nonumber \\
& +\tfrac{96}{125}\,W^\gamma{}_{\mu\nu\beta,\gamma\alpha}\,
W^{\rho\mu\nu\beta}{}_\rho{}^\alpha +\tfrac{74}{25}\,
W_\beta{}^{\alpha\gamma\mu}\,W_{\nu\alpha\gamma\mu}\,
W^\beta{}_{\rho\sigma}{}^{\nu,\sigma\rho} \,
\nonumber \\
& + \tfrac{208}{5}\,W_{\mu\beta\gamma\alpha}\,W_\sigma{}^{\nu\rho\alpha}\,
W^\mu{}_{\nu\rho}{}^{\sigma,\gamma\beta} 
-8\,W_\alpha{}^\gamma{}_\beta{}^\mu\,W^\alpha{}_\nu{}^\beta{}_\rho\,
W^\nu{}_\gamma{}^\rho{}_{\mu,\sigma}{}^\sigma \,
\nonumber \\
& +\tfrac{16}{5}\,W_{\alpha\gamma\mu\rho}\,
W_{\beta\nu}{}^{\alpha\gamma}\,W^{\beta\nu\mu\rho}{}_\sigma{}^\sigma
-\tfrac{144}{25}\,W^\gamma{}_\alpha{}^\mu{}_\beta\,
W_{\rho\gamma}{}^\nu{}_\mu{}^\rho\,W_\sigma{}^\alpha{}_\nu{}^{\beta,\sigma} \,
\nonumber \\
& +\tfrac{104}{5}\,
W_\alpha{}^\gamma{}_\beta{}^\mu\,W^\beta{}_\mu{}^{\sigma\nu,\alpha}\,
W_{\rho\gamma\sigma\nu}{}^\rho
-\tfrac{88}{25}\,W_{\alpha\beta\gamma\mu}\,
W_{\rho\nu}{}^{\alpha\beta,\rho}\,W_\sigma{}^{\nu\gamma\mu,\sigma}\;,
\\
I_2 =\;& W_\beta{}^{\alpha\gamma\mu}\,W_{\nu\alpha\gamma\mu}\,
W^\beta{}_{\rho\sigma}{}^{\nu,\sigma\rho}
+5\,W_{\alpha\gamma\mu\rho}\,W_{\beta\nu}{}^{\alpha\gamma}\,
W^{\beta\nu\mu\rho}{}_\sigma{}^\sigma \,
\nonumber \\
& +5\,W_{\alpha\beta\gamma\mu}\,W^{\alpha\beta\rho\sigma}{}_\nu\,
W^{\gamma\mu}{}_{\rho\sigma}{}^\nu
+\tfrac{12}{5}\,W_{\alpha\beta\gamma\mu}\,W_{\rho\nu}{}^{\alpha\beta,\rho}\,
W_\sigma{}^{\nu\gamma\mu,\sigma}\;,
\\
I_3 =\;& W_\beta{}^{\alpha\gamma\mu}\,W_{\nu\alpha\gamma\mu}\,
W^\beta{}_{\rho\sigma}{}^{\nu,\sigma\rho}
-20\,W_\alpha{}^\gamma{}_\beta{}^\mu\,W^\alpha{}_\nu{}^\beta{}_\rho\,
W^\nu{}_\gamma{}^\rho{}_{\mu,\sigma}{}^\sigma \,
\nonumber \\
& -\tfrac{48}{5}\,W^\gamma{}_\alpha{}^\mu{}_\beta\,
W_{\rho\gamma}{}^\nu{}_\mu{}^\rho\,W_\sigma{}^\alpha{}_\nu{}^{\beta,\sigma}
-20\,W^\alpha{}_\mu{}^\gamma{}_\beta\,W^\mu{}_\rho{}^\beta{}_{\sigma,\nu}\,
W^\rho{}_\alpha{}^\sigma{}_\gamma{}^\nu\;.
\end{align} 
Therefore, the most general action for eight-dimensional conformal gravity 
is an arbitrary linear combination of the previous $3+7$ invariants:
\begin{equation}
S_8[g_{\mu\nu}] = \int\diff^8 x\,
(w_1\,{\cal I}_1+w_2\,{\cal I}_2+w_3\,{\cal I}_3+w_6\,{\cal I}_6+\dots 
+ w_{12}\,{\cal I}_{12})\;.
\end{equation}
We compute the variation of $S_8[g_{\mu\nu}]$ with respect to the metric 
and impose that it should vanish for an Einstein metric, to find that the 
most general action compatible with an Einstein sector is the following 
unique one, up to an overall factor:
\begin{equation}\label{8dConformalAction}
\tilde{S}_8 = -\tfrac{1}{4}\int_{M_8}\text{d}^8 x\,
({\cal I}_1 - \tfrac{23}{25}\,{\cal I}_2 -\tfrac{21}{25}\,{\cal I}_3 
- 39\,{\cal I}_6 + \tfrac{2}{5}\,{\cal I}_7 - \tfrac{7}{2}\,{\cal I}_8 
+ \tfrac{124}{5}\,{\cal I}_9 + \tfrac{9}{20}\,{\cal I}_{10} 
-4\,{\cal I}_{11} - \tfrac{568}{5}\,{\cal I}_{12})\,.
\end{equation}
To reach this result, we used an algebraic method very similar to the one 
described in \cite{Boulanger:2004zf}. 
The challenge is to properly take into account the algebraic and differential 
Bianchi identities on the $W$-tensors. 
After variation, we imposed the Einstein manifold condition, which requires the 
Schouten tensor appearing in the expression of the $W$-tensors to be 
proportional to the metric. We then selected a basis of independent structures, 
so that the whole variation written in the basis identically vanishes if and 
only if all the coefficients (which are linear combination of the $w_i$ constants)  
vanish. The computations, although relatively straightforward, are very tedious 
and could not have been done without the use of Mathematica and the suite
of packages xAct \cite{martin2020xact}, 
including xTras \cite{nutma2014xtras}.

\paragraph{Relation with the $Q$-curvature.}

The eight-dimensional $Q$-curvature was computed in 
\cite{Gover:2002ay}. 
We verified that there exists a suitable boundary term $\mathcal{V}^\alpha$ such that
\begin{align}
{\cal Q}_8 =\,& \partial_\alpha\,{\cal V}^\alpha - \tfrac{1}{128}\,{\cal E}_8(R) 
\\
&- \tfrac{5}{3}\,(\mathcal{I}_1 - \tfrac{23}{25}\,\mathcal{I}_2 -\tfrac{21}{25}\,\mathcal{I}_3 
- 39\,\mathcal{I}_6 + \tfrac{2}{5}\,\mathcal{I}_7 - \tfrac{7}{2}\,\mathcal{I}_8 + \tfrac{124}{5}\,\mathcal{I}_9 
+ \tfrac{9}{20}\,\mathcal{I}_{10} -4\,\mathcal{I}_{11} - \tfrac{568}{5}\,\mathcal{I}_{12})\;,
\nonumber
\end{align}
where ${\cal E}_8(R)$ is the Euler density
\begin{equation}
{\cal E}_8(R) = 
\sqrt{|g|}\,
\varepsilon_{\mu\nu\rho\sigma\kappa\lambda\xi\phi}\,
\varepsilon^{\alpha\beta\gamma\delta\varepsilon\zeta\theta\iota}\,
R^{\mu\nu}{}_{\alpha\beta}\,R^{\rho\sigma}{}_{\gamma\delta}\,
R^{\kappa\lambda}{}_{\varepsilon\zeta}\,R^{\xi\phi}{}_{\theta\iota}\;,
\end{equation}
and the last term is proportional to the integrand of the eight-dimensional 
conformal action with Einstein sector, see \eqref{8dConformalAction}.
Then, integrating on the closed manifold $M_8$ with Euler characteristic 
$\chi(M_8)\,$, we find 
\begin{equation}\label{S8andQ8}
\tilde{S}_8[g_{\mu\nu}] = \tfrac{576 \pi^4}{5}\,\chi(M_8)
+ \tfrac{3}{20}\,\int_{M_8}\diff^8 x\,\,{\cal Q}_8 \;,
\end{equation}
and the variational derivative 
$O^{(8)}_{\mu\nu}=\frac{1}{\sqrt{|g|}}\frac{\delta {\cal Q}_8}{\delta g^{\mu\nu}}$ is 
a symmetric rank-two tensor, divergenceless, traceless, of conformal dimension 
$-6$, that vanishes on Einstein manifolds.

\section{Discussion}
\label{sec:Discu}

The $Q$-curvature can be defined 
and explicitly constructed via the ambient method of 
\cite{FeffermanGraham2002,graham2005ambient,graham2007holographic};  
see also \cite{chang2008q} for a review, as well as the book \cite{juhl2009}.
That the functional derivative (with respect to the inverse metric) 
of the integrated $Q$-curvature on an even-dimensional closed manifold 
is proportional to the obstruction tensor was already proven in 
\cite{graham2005ambient}, see Theorem 1.1 therein. Moreover, 
in the Theorem 2.1 of the latter work, the obstruction tensor 
was shown to obey four defining properties. To paraphrase their 
Theorem 2.1, if $(M^n, [g])$ is a conformal manifold of even dimension 
$n=2m \ge 4$, then there exists a natural symmetric $2$-tensor 
${O}_{\mu\nu}$, called the \emph{ambient obstruction tensor}, 
with the following properties:
\begin{enumerate}
  \item \textbf{Naturality:} ${O}_{\mu\nu}$ is natural, i.e., it can be expressed as a universal polynomial in the metric, its inverse, the curvature, and covariant derivatives of the curvature.
  
  \item \textbf{Linearisation:} ${O}_{\mu\nu}$ is a symmetric,
  conformally covariant tensor whose expression starts with
  \begin{equation}\label{symbol}
    {O}_{\mu\nu} 
    = \frac{1}{3-n}\bigtriangleup^{m-2} \left( \nabla^\alpha \nabla^\beta W_{\mu\alpha\nu\beta} \right) + \text{lower order terms},
  \end{equation}
  where $\bigtriangleup = g^{\alpha\beta} \nabla_\alpha \nabla_\beta$ 
  (or the D'Alembertian, in Lorentzian signature), 
  and ``lower order terms'' involve terms with fewer derivatives 
  of the curvature.
  The tensor ${O}_{\mu\nu}$ has conformal weight $2-n$, meaning that 
  for $\widehat{g} = e^{2\omega} g\,$,
  $\widehat{{O}}_{\mu\nu}= e^{(2-n)\omega}\,{O}_{\mu\nu}\;$.
  
  \item \textbf{Trace and divergence:} Traceless and divergence-free: $g^{\mu\nu} {O}_{\mu\nu} = 0 =\nabla^\mu {O}_{\mu\nu}$.
  
  \item \textbf{Vanishing on conformally Einstein metrics:} If $g$ is conformal to an Einstein metric, then
  \begin{equation}
    {O}_{\mu\nu} = 0\;.
  \end{equation}
\end{enumerate}
As far as we understand, it is unknown whether these four 
properties uniquely specify the obstruction tensor; it seems likely 
that they do.

For possible axiomatic uniqueness results concerning 
the obstruction tensor, 
another line of investigation relies on group representation theory 
and the Bernstein-Gelfand-Gelfand (BGG) machinery 
as developed for curved geometries in \cite{CapSlovakSoucek2001}
with detour complexes, 
and in particular its application \cite{gover2006ambient,Branson:2006ug}
to the (curved) conformal case,  
where the obstruction tensor appears as a canonical object 
whose vanishing allows sequences of differential operators 
to be conformally invariant and to form a complex. 
Still, in this curved case set up, 
we are not aware of the way to translate in group-theoretical 
terms the condition that the obstruction tensor should vanish 
on Einstein manifolds, neither are we aware of proofs of 
uniqueness for the obstruction tensor.

From the perspective of the present work and in the particular dimensions
six and eight, the uniqueness of the action, and hence, of the obstruction 
tensor, appears from the requirement that the latter should vanish 
on an Einstein metric. For that result, it appeared crucial that 
the first pointwise invariant, ${\cal I}_1\,$, should appear in the 
Lagrangian density for conformal gravity, and 
it is the property \eqref{symbol} that encodes the fact that the invariant
${\cal I}_1\,$ should appear in the Lagrangian density.
It is tempting to conjecture that, in arbitrary even dimension, 
the obstruction tensor is uniquely specified by the four conditions 
given in \cite{graham2005ambient}, and therefore that the conformal 
gravity action that possesses an Einstein sector is unique up to 
boundary terms, given by the integrated $Q$-curvature in 
dimension $2m$.

In this work, we have reviewed and re-derived the conformal actions 
in dimensions 4 and 6 that admit an Einstein sector, 
and we have compared them with the $Q$-curvature in the corresponding dimensions. 
We have then explicitly built the conformal gravity action  
in eight dimensions that admits an Einstein sector, and have shown that 
it is unique, up to boundary terms and overall normalisation. 
We have compared it with the integral of the $Q$-curvature in eight 
dimensions and found that they are proportional, up to an additive constant 
related to the Euler characteristic of the manifold, and up to boundary terms. 
In arbitrary even dimension, we conjecture 
that the conformal gravity theory that admits an Einstein sector is unique. 
It can be defined by the integral of the $Q$-curvature on the 
even-dimensional spacetime manifold.

In a forthcoming work in collaboration with Giorgos Anastasiou, 
Ignacio J. Araya, and Rodrigo Olea, we will further study the 
eight-conformal gravity action we have built in this work. 
We plan to discuss its physical properties in relation to holography and 
the volume-renormalisation of 8D Einstein theory, in particular addressing 
the problem of the conformal completion of 8D Hilbert-Einstein action, 
along the lines of 
\cite{Anastasiou:2018mfk,Anastasiou:2020mik,Anastasiou:2020zwc,Anastasiou:2024rxe}\footnote{See \cite{Jia:2021hgy} for a related work.}.

\section*{Acknowledgements}

We are very grateful to Thomas Basile and Jeffrey S. Case for helping
us to clarify our views on the Fefferman-Graham obstruction tensor. 
It was a pleasure to discuss with Giorgos Anastasiou, Ignacio J. Araya, 
and Rodrigo Olea. 
We also benefited from discussions with Xavier Bekaert,  
Rod Gover, Yannick Herfray, and Rodrigo Olea during the workshop 
\href{https://web.umons.ac.be/pucg/en/event/workshop-conformal-higher-spins-twistors-and-boundary-calculus/}{\tt Conformal higher spins, twistors and 
boundary calculus} that took place at UMONS on 30 June -- 4 July 2025,  
where one of us (D.~R.~) presented part of the present results. 


\providecommand{\href}[2]{#2}\begingroup\raggedright\endgroup

\end{document}